# Solid-state Mamyshev oscillator

MINGMING NIE,[1,2] JIARONG WANG,[1] AND SHU-WEI HUANG[1,3]

[1]*Department of Electrical, Computer & Energy Engineering, University of Colorado Boulder, Boulder, CO 80309, USA*
[2]*Mingming.Nie@colorado.edu*
[3]*ShuWei.Huang@colorado.edu*

**Abstract:** We present the first design and analysis of a solid-state Mamyshev oscillator. We utilize the phase-mismatched cascaded quadratic nonlinear process in periodically poled lithium niobite waveguide to generate substantial spectral broadening for Mamyshev mode-locking. The extensive spectral broadening bridges the two narrowband gain media in the two arms of the same cavity, leading to a broadband mode-locking not attainable with either gain medium alone. Two pulses are coupled out of the cavity and each of the output pulses carries a pulse energy of 25.3 nJ at a repetition rate of 100 MHz. The 10-dB bandwidth of 2.1 THz supports a transform limited pulse duration of 322 fs, more than 5 times shorter than what can be achieved with either gain medium alone. Finally, effects of group velocity mismatch, group velocity dispersion, and nonlinear saturation on the performance of Mamyshev mode-locking are numerically discussed in detail.

## 1. Introduction

In the past decades, ultrafast lasers attract much attention owing to the widespread scientific and industrial applications [1-3]. Much effort has been made to pursue mode-locked pulses with ever larger pulse energy, shorter pulse duration and higher stability [4-6]. Recently a new mode-locked fiber laser design, Mamyshev oscillator, has been revisited and superior performance has been demonstrated [7, 8]. To sustain the mode locking status in Mamyshev oscillator, the necessary ingredients are the spectral broadening (or shifting), two spectrally shifted bandpass filters, and high laser gain to compensate for the filter loss. The mode-locking mechanism of Mamyshev oscillator can be explained by the different transfer functions for pulses with different intensities. Low intensity pulses producing insufficient self-phase modulation would be blocked by the offset filter, while the high intensity ones would go through, resulting in an effective saturate absorber (SA) [9].

Compared to the all-normal-dispersion (ANDi) fiber laser [10], which is fundamentally limited by nonlinear phase accumulation at ~$10\pi$, the Mamyshev oscillator delivers order-of-magnitude improvements in both the peak nonlinear phase (~$60\pi$) and the peak power in an environmentally stable design [8]. The outstanding performance is attributed to the evident noise-suppressing by the effective SA. To date, pulse energy of >1 µJ, compressed pulse duration of 41 fs thus peak power of >10 MW has been realized by the employ of large-mode-area photonic crystal fiber [11]. These capabilities make fiber Mamyshev oscillators competitive with the typical Ti:sapphire oscillators, opening a door to achieve the above-mentioned goals for the ultrafast lasers. The main limitation of fiber Mamyshev oscillator is that the operation in the anomalous group-delay dispersion (GDD) regime is prohibitive due to the high energy pulse breakup problem. Fiber Mamyshev oscillators to date work either at 1 µm where the normal GDD fiber is readily available [8, 11-13] or at 1.5 µm where specialty fibers can be utilized to shift the GDD [14]. Operation at longer wavelengths with Tm-doped silica fiber and Er-doped ZBLAN fiber, characteristic of large anomalous GDD and Kerr nonlinearity, thus poses a nontrivial challenge [9].

In this paper, we present the first design and analysis of a solid-state Mamyshev oscillator. To date, the concept of Mamyshev oscillator has not been applied to solid-state lasers due to the lack of proper mechanism for achieving substantial spectral broadening. We address the challenge by utilizing the phase-mismatched cascaded quadratic nonlinear process [15-18] to generate enough spectral broadening for Mamyshev mode-locking. In particular, we propose to use a periodically poled lithium niobite (PPLN) waveguide that can provide an effective Kerr nonlinear coefficient two orders of magnitude higher than that in optical fibers. Large nonlinear phase can be quickly accumulated in a cm-long PPLN waveguide, making possible the compact solid-state Mamyshev oscillator. Importantly, the sign and magnitude of the effective Kerr nonlinearity can be controlled by the design and tuning of the periodically poled structure. Such flexibility mitigates the requirement of large normal GDD that limits the application of Mamyshev oscillator in longer wavelengths. The lower GDD of solid-state Mamyshev oscillators also have the added benefit of better phase noise and time jitter performance [19].

In addition, we investigate the use of two narrowband gain media with distinct gain spectra in the two arms of the same cavity. The extensive spectral broadening in the PPLN waveguide bridges the otherwise spectrally separated gain media and a 10-dB bandwidth of 2.1 THz can be obtained. Two pulses are coupled out of the cavity and each of the output pulses carries a pulse energy of 25.3 nJ at a repetition rate of 100 MHz. While a simple GDD compensation can already compress the output pulse down to 373 fs, a finer dispersion compensation can further compress the pulse to its transform limit of 322 fs that is more than 5 times shorter than what can be achieved with either gain medium alone. Finally, influences of group velocity mismatch (GVM), group velocity dispersion (GVD), and nonlinear saturation on the performance of Mamyshev mode-locking are discussed in detail.

## 2. Simulation setup

Without loss of generality, we consider a unidirectional ring laser cavity (as shown in Fig. 1) with two bulk gain medium, a-cut Nd:YVO$_4$ and a-cut Nd:GdVO$_4$. Within the slowly varying envelope approximation, the electric field $A(z,\tau)$ propagating through the gain crystal is given by the generalized nonlinear Schrödinger equation (NLSE)

$$\frac{\partial A}{\partial z} = \frac{g}{2}A + \left(\frac{g}{2\Omega_g^2} - j\frac{\beta_2}{2}\right)\frac{\partial^2 A}{\partial \tau^2} + j\gamma |A|^2 A , \quad (1)$$

where $z$ is propagation distance in the gain crystal, $\tau$ is the local time in the moving frame, $\gamma$ is the Kerr nonlinear coefficient, $\beta_2$ is the GVD coefficient, $g$ is the intensity gain coefficient, and $\Omega_g$ is the gain bandwidth. The intensity gain coefficient $g$ is defined as

$$g = g_0(z) \bigg/ \left(1 + \frac{\int |A(z)|^2 d\tau}{P_{sat}T_r}\right), \quad (2)$$

where $g_0$ is the small signal gain, $P_{sat}$ is the saturation power, and $T_r$ is the repetition period.

Both gain crystals are 20 mm long and the beam diameters within the crystals are set at 200 µm to minimize the nonlinearity. Small signal gain $g_0$ and saturation power $P_{sat}$ of both gain crystals are set at 44 dB [20, 21] and 1.46 W, respectively. Center emission wavelengths of Nd:YVO$_4$ and Nd:GdVO$_4$ crystals are 1064.3 nm [22] and 1062.9 nm [21], respectively. Both gain crystals have a full-width-half-maximum (FWHM) gain bandwidths $\Omega_g$ of 1.0 nm [23, 24], supporting a Gaussian transform-limited pulse duration of 1.7 ps [25, 26].

The two band-pass filters (BPFs) both have super-Gaussian transmission profiles:

$$T_\pm(\nu) = \exp\left[-\left(\frac{\nu - (\nu_0 \pm \Delta\nu)}{\sigma/2\sqrt{\ln 2}}\right)^8\right], \quad (3)$$

where $\nu_0$ =281.87 THz (1063.6 nm) is the center frequency of the two BPFs, $\sigma$ =0.15 THz (0.55 nm) is the BPF bandwidth, and $\Delta\nu$ =0.19 THz (0.7 nm) is half of the offset between the two BPFs that is essential for the Mamyshev mode-locking.

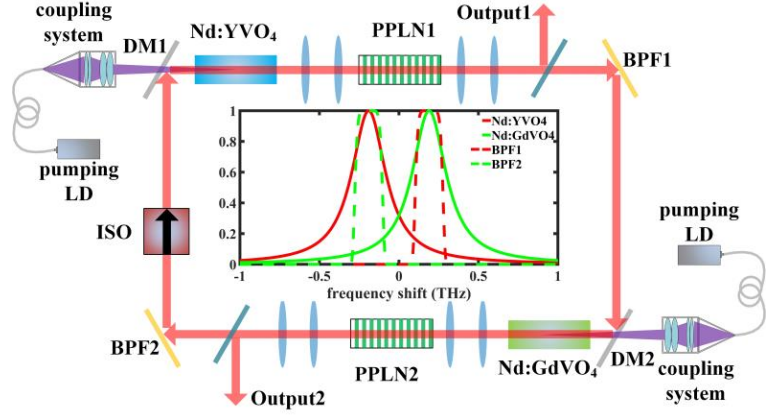

Fig. 1. Setup of the solid-state Mamyshev oscillator. DM: dichroic mirror; ISO: isolator. Inset: gain curves of the two media and transmission profiles of the two BPFs.

Slightly phase-mismatched PPLN ridge waveguides are introduced to generate the sufficient spectral broadening for the mode locking. The waveguide has a dimension of 15 mm (length) × 11.5 µm (width) × 10 µm (height). Fig. 2(a) shows the numerically simulated mode profile of the fundamental transverse-electric ($TE_{00}$) mode at the FF wavelength, with the effective mode area being 85 µm². The $TE$-polarized modes utilize the highest second-order nonlinear tensor component, resulting in a large effective second-order nonlinear coefficient $d_{eff}$ =14 pm/V. Phase mismatch is set at $\Delta k$ =±4π mm$^{-1}$ such that an effective Kerr nonlinear coefficient $\gamma_{eff}$ [15] of ±500 m$^{-1}$kW$^{-1}$ can be achieved. The large and controllable $\gamma_{eff}$ ensures strong spectral broadening for the proper pulse mode-locking.

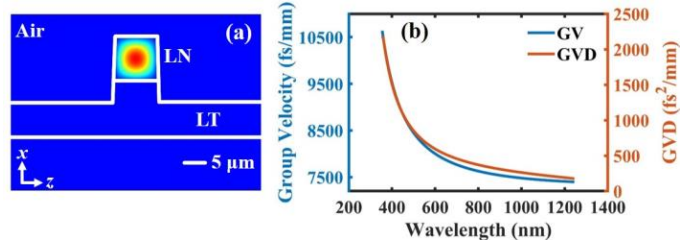

Fig. 2. Details of the PPLN ridge waveguide. (a) Mode profile ($E_z$ component) of the fundamental $TE_{00}$ mode at 1064 nm. LN: Lithium niobate; LT: Lithium tantalate; (b) GV and GVD coefficients as a function of wavelength.

Within the slowly varying envelope approximation, the coupled propagation equations governing the interaction between the FF and SH in the PPLN waveguide are given by

$$\frac{\partial A_{FF}}{\partial z} + j\frac{\beta_{2FF}}{2}\frac{\partial^2 A_{FF}}{\partial \tau^2} = j\frac{\omega_{FF} d_{eff}}{n_{FF} c} A_{SH} A_{FF}^* e^{-j\Delta kz} \quad (4)$$

$$\frac{\partial A_{SH}}{\partial z} + GVM\frac{\partial A_{SH}}{\partial \tau} + j\frac{\beta_{2SH}}{2}\frac{\partial^2 A_{SH}}{\partial \tau^2} = j\frac{\omega_{SH} d_{eff}}{2n_{SH} c} A_{FF}^2 e^{j\Delta kz} \quad (5)$$

where $A$ is the electric field, $GVM$ is the group velocity (GV) mismatch between SH and FF fields, $\beta_2$ is the GVD coefficient, $\omega$ is the optical center frequency, and $n$ is the linear refractive index. Figure 2(b) plots the dispersion curves, including the GV and the GVD, which are numerically simulated and entered into the coupled propagation equations. $\Delta k = 2k_1 - k_2 + \vec{k}_\Lambda$ is the wave-vector mismatch, where $\vec{k}_\Lambda$ is the wave-vector induced by the periodical poling structure. The sign and magnitude of $\Delta k$ determine the nonlinear phase shift and effective nonlinearity coefficient, and it can be controlled by the design and tuning of the periodical poling structure.

## 3. Numerical results

In the first design, the cavity length is set at 1.5 m such that the fundamental mode-locked repetition rate is 200 MHz, $\Delta k$ for PPLN1 is $4\pi$ mm$^{-1}$ and for PPLN2 is $-4\pi$ mm$^{-1}$. A Gaussian pulse with an energy of 11 pJ and a duration of 5 ps is used as the initial condition in the simulation to ignite the mode-locking process. Such seed pulse is comparable to what has been used in the fiber platform [8] and it can be easily generated by electronically modulating laser diode through gain switching [27] or electro-optic modulation [28, 29]. By solving the Eqs. (1), (4) and (5) with the standard split-step Fourier method, a stable pulse as depicted in Fig. 3 can sustain in the cavity just after 6 roundtrips. Two pulses are coupled out of the cavity and each of the output pulses carries a pulse energy of 15.3 nJ. The optical spectrum is heavily broadened through the cascaded quadratic nonlinear process in the PPLN waveguides, with the spectral asymmetry resulting from the large GVM between FF and SH. Details will be discussed in the next section. The output pulse is highly chirped with a FWHM duration of 6.9 ps. Its self-steepening at the trailing edge is attributed to the temporal walk-off in the cascaded quadratic nonlinear process and details will also be discussed in the next section. As $\Delta k$ in the two PPLN waveguides are chosen to have an opposite sign, the two output spectra and pulse chirps are mirror images of each other. Nevertheless, both output pulses can be well compressed down to 512 fs with only GDD compensation.

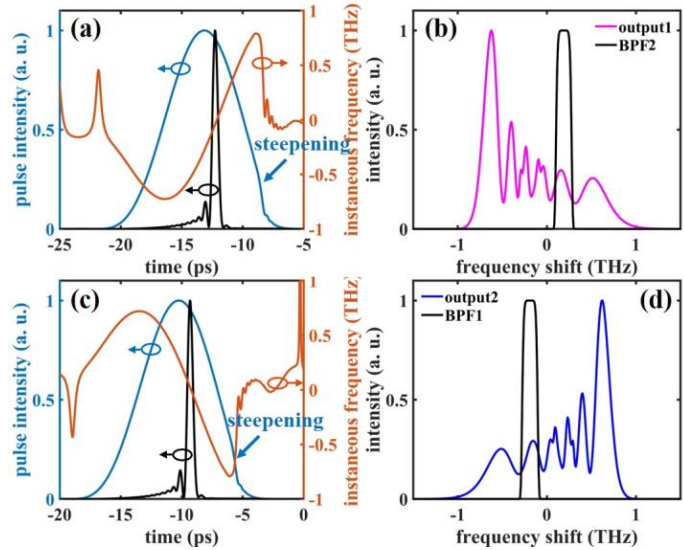

Fig. 3. Performance of the dual outputs, (a)(b) for output1 and (c)(d) for output2, from the solid-state Mamyshev oscillator. (a)(c): normalized pulse profiles before (blue curve) and after (black curve) GDD compensation; (b)(d): normalized output spectra, overlaid with the BPF transmission profiles.

Figure 4 shows the intra-cavity evolution of the pulse shape and pulse spectrum. Clearly the band-pass filters play the key role in shaping the intra-cavity pulse and promoting the mode locking physics. The pulse duration right after the two filters reaches its minimum of 4.8 ps with reduced chirp, while it continues to grow to 5.8 ps and 6.9 ps propagating through the gain crystals and the PPLN waveguides. Spectrally, the band-pass filters select only a small portion of the optical spectrum highly broadened in the PPLN waveguides by the cascaded quadratic nonlinear process. In contrast, the nonlinear phase shift in the gain crystals is negligible at 0.56 mrad and thus gain narrowing instead of spectral broadening is observed. Of note, the filter bandwidth has to be chosen carefully so that it is narrow enough to transmit only one spectral ripple in the broadened spectrum (Fig. 3(b) and 3(d)). When the filter bandwidth is set too broad, the transmitted pulse shape will have strong oscillating structures, leading to inconsistent spectral broadening and unstable mode locking.

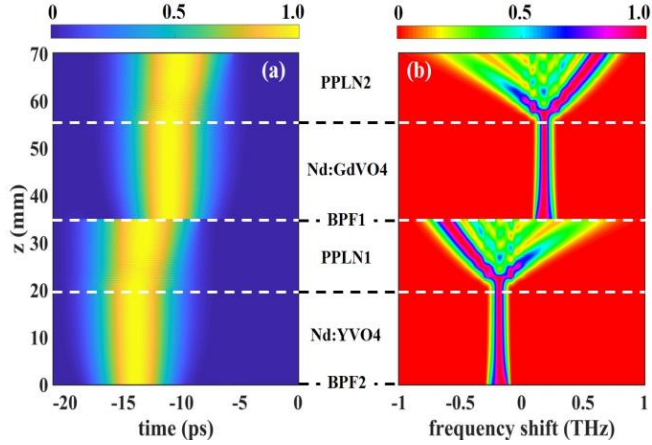

Fig. 4. Intra-cavity evolution of pulse profiles (a) and optical spectra (b), normalized at each step to show the details of the profiles.

To achieve even higher pulse energy, one can either reduce the fundamental mode-locked repetition rate or increase the beam diameter within the gain crystal. Figure 5 shows the output pulse profile and optical spectrum when the fundamental mode-locked repetition rate is reduced to 100 MHz, while all the other parameters are kept unchanged. Pulse energy of 25.3 nJ and 10-dB bandwidth of 2.1 THz is obtained. While a simple GDD compensation can already compress the output pulse down to 373 fs, a finer dispersion compensation can further compress the pulse to its transform limit of 322 fs that is more than 5 times shorter than what can be achieved with either gain medium alone.

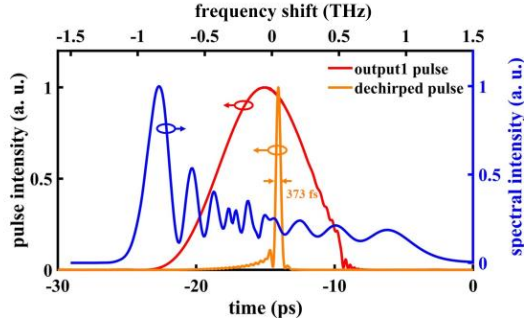

Fig. 5. Output pulse profile and optical spectrum of the second design.

## 4. Discussion

### 4.1 Influence of GVM on spectral broadening

In the PPLN waveguides, the FF fields acquire Kerr-like nonlinearities from conversion and back-conversion between the FF and SH fields [30]. The FF-SH interaction marks the fundamental difference between the pure Kerr and Kerr-like nonlinearities, where GVM and nonlinear saturation play significant roles in the dynamics of spectral broadening and pulse shaping. The GVM effect manifests itself in the temporal walk-off between the FF and the SH pulses, the splitting of the SH pulse, and the self-steepening (SS) of the FF pulse's trailing edge (Fig. 6). The GVM-induced SS [31] then leads to the spectral asymmetry (Fig. 4(b)), with the sign of $GVM \cdot \Delta k$ determining whether the spectrum is effectively blue-shifted or red-shifted. As demonstrated in Section 3, such asymmetric spectral broadening can be utilized to better bridge gain crystals with well separated gain spectra and achieve a broadband mode-locking not attainable with either gain crystal alone.

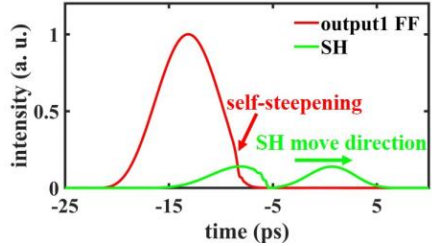

Fig. 6. FF and SH pulse profiles, illustrating the GVM effect.

To further elucidate the GVM effect on the spectral broadening, we launch a 2-ps FF pulse into a 40-mm-long PPLN waveguide and record the spectral evolution of both FF and SH fields (Fig. 7). Except for the length, all other PPLN waveguide parameters are identical to the ones used in Section 3. Due to the GVM, two distinct stages can be identified in the evolution of cascaded quadratic nonlinear process. Temporally, the two evolution stages correlates to the SH pulse splitting as shown in Fig. 6. In the first stage, intensive cycles of conversion to back-conversion occurs as expected and such cascaded quadratic nonlinear process results in rapid spectral broadening (Fig. 7(a) and 7(b)). Even though the magnitude of the spectral broadening is greater on the red detuned side, the optical power spectral density on the blue detuned side is however much higher, resulting in an effective blue-shift in the broadened optical spectrum. Correspondingly, blue-shifted SH are generated in the process (Fig. 7(c) and 7(d)). Of note, the sign of $GVM \cdot \Delta k$ determines whether such effective SS-induced spectral shift is towards the blue or red sides. Additionally, GVM leads to imperfect back-conversion as the SH pulse will experience a temporal walk-off from the FF pulse. Such GVM-induced loss [31] will lead to a gradual slowdown of the spectral broadening. In the second stage, the spectral broadening on the blue detuned side continues to grow despite apparent slowdown as expected. In contrast, the spectral broadening on the red detuned side is nearly stopped and it is accompanied by the growth of a highly redshifted narrowband SH peak. Similar to the generation of soliton Cherenkov radiation [32], the phase-matched frequency component of the SH will grow monotonically along the propagation due to constructive interference. The phase-matching condition can be revised from [33] to take into account the spectral blue-shift:

$$\omega_{SH\_R} = \Delta\omega_{SH} - \Delta k/GVM \ , \qquad (6)$$

where $\Delta\omega_{SH}$ is the SH blue-shift in the first stage. Figure 7(e) plots the resonant SH frequencies obtained numerically (solid line) and calculated analytically (dashed line), showing a good agreement. The resonant SH depletes the red detuned edge of the FF spectrum and effectively limits its further spectral broadening. The frequency shift limit imposed by such GVM-induced depletion, according to Eq. (6), is thus proportional to the wave-vector mismatch and inverse proportional to the GVM.

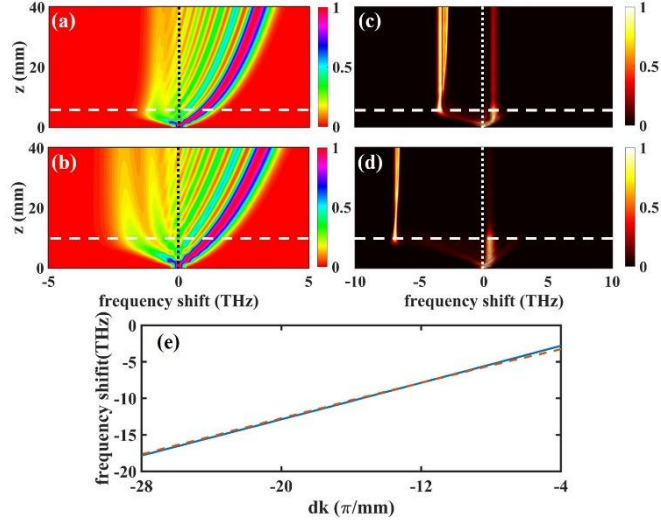

Fig. 7. (a)-(d): Evolution of FF and SH optical spectra with the propagating distance, normalized at each step to show the details of the profiles. The peak power is 15 kW and the GVM is 800 fs/mm. $\Delta k$ used in (a)(c) and (b)(d) are $-4\pi$ mm$^{-1}$ and $-10\pi$ mm$^{-1}$, respectively. (e) Resonant SH frequencies obtained numerically (solid line) and calculated analytically (dashed line).

Figure 8(a) plots a few snapshots of the FF and SH optical spectra along the propagation. Two spectral peaks, FF_blue and FF_red, are marked and tracked to quantitatively study the spectral broadening. Figure 8(b) summarizes the peak frequency shifts as a function of wave-vector mismatch. Importantly, because of the GVM, the blue and red shifts reach their maximum values at different wave-vector mismatches. FF_blue reaches the maximum shift when $\Delta k = -9\pi$ mm$^{-1}$ and it is an intricate balance between minimizing the GVM-induced loss that favors larger $|\Delta k|$ and maximizing the effective Kerr nonlinearity that prefers lower $|\Delta k|$. The same balance applies to FF_red, with another factor GVM-induced depletion that favors larger $|\Delta k|$ coming into play. Thus the optimal wave-vector mismatch for the FF_red is shifted to a larger value when $\Delta k = -26\pi$ mm$^{-1}$.

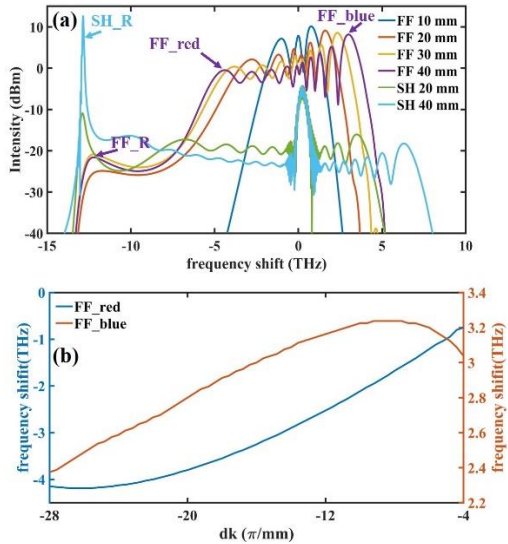

Fig. 8. (a) Snapshots of FF and SH optical spectra along the propagation, with set Δ$k$ at -20π mm$^{-1}$. Resonant frequencies are labeled as FF_R and SH_R. (b) Peak frequency shifts of FF_red and FF_blue as a function of wave-vector mismatch.

*4.2 Influence of nonlinear saturation on pulse shaping*

With the GVD taken into account, both FF and SH fields would experience pulse compression and eventually pulse breakup in the PPLN waveguide when $GVD·\Delta k$ is negative. The pulse breakup problem then imposes an upper limit on the pulse peak power to 10 kW in the absence of GVM. When the GVM is considered, pulse compression greatly slows down due to the GVM-induced loss and the pulse peak power can be increased to 200 kW without any pulse breakup. The phenomena resemble the behavior in standard nonlinear fiber optics [34].

On the other hand, Fig. 9 shows a unique dynamics that pertains to the cascaded quadratic nonlinearity. At a high peak power of 15 kW and a small wave-vector mismatch of -10π mm$^{-1}$, both FF and SH pulses are distorted by dense ripples even in the absence of GVM and GVD of FF (see Fig. 9(a)). This can be understood by the nonlinear saturation, not present in standard nonlinear fiber optics, which dictates the conversion to back-conversion cycle to be no longer a constant but a function of pulse intensity [35]. Figure 9(b) shows the case when $GVD·\Delta k$ is positive and both FF and SH fields experience pulse broadening without the influence of GVM. Pulse distortion by ripples remain even though the oscillation period is proportionally increased with the amount of the pulse broadening. The emergence of such ripples limits the pulse compressibility at the end.

Interestingly, when the GVM effect is considered, the ripples can either be confined to one side of the pulse (Fig. 9(c)) or completely suppressed (Fig. 9(d)). Furthermore, we find that pulse compression regime ($GVD·\Delta k$ <0) is counterintuitively favored for achieving a clean pulse shape and a large spectral broadening simultaneously. Even though the exact mechanism awaits more detailed investigation, Fig. 9 illustrates the efficacy of GVM to counteract the nonlinear saturation in the pulse compression regime and stabilize the mode-locking at a higher energy level.

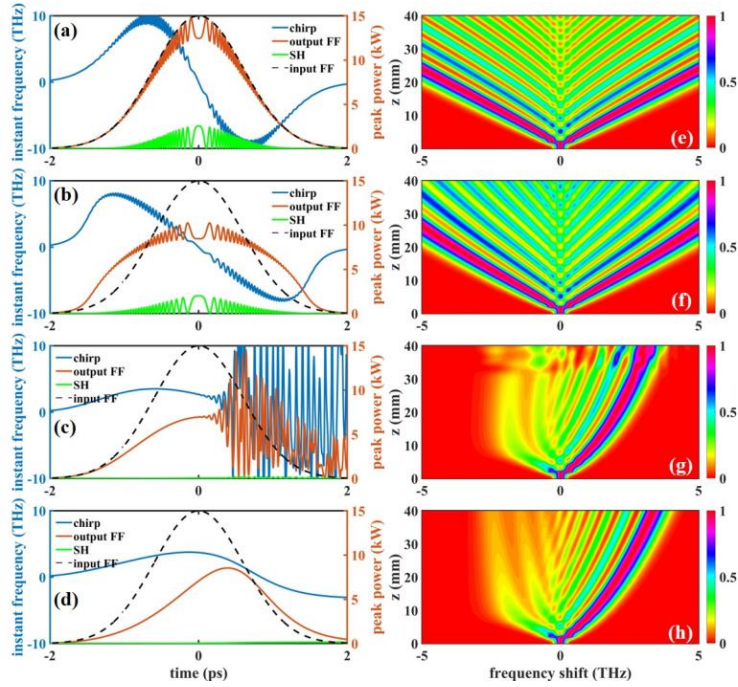

Fig. 9. (a) ~ (d): FF and SH pulse profiles and the optical chirp of FF. (e) ~ (h): evolution of FF optical spectrum with the propagating distance, normalized at each step to show the details

of the profiles. Δ*k* is set at -10π mm$^{-1}$ and the peak power is set at 15 kW. (a) and (e): GVM = 0, GVD = 0 for FF; (b) and (f): GVM = 0, GVD = -237 fs$^2$/mm; (c) and (g): GVM = 800 fs/mm, GVD = -237 fs$^2$/mm; (d) and (h): GVM = 800 fs/mm, GVD = 237 fs$^2$/mm.

## 5. Conclusion

In conclusion, we present the first design and analysis of a solid-state Mamyshev oscillator. We utilize the phase-mismatched cascaded quadratic nonlinear process in PPLN waveguides to achieve substantial spectral broadening that bridges the otherwise spectrally separated gain media. Two pulses are coupled out of the cavity and each of the output pulses carries a pulse energy of 25.3 nJ at a repetition rate of 100 MHz. A 10-dB bandwidth of 2.1 THz is obtained and a simple GDD compensation can already compress the output pulse down to 373 fs. A finer dispersion compensation can further compress the pulse to its transform limit of 322 fs that is more than 5 times shorter than what can be achieved with either gain medium alone.

Furthermore, the influences of GVM, GVD, and nonlinear saturation on the performance of Mamyshev mode-locking are discussed and summarized: 1, GVM would alleviate the pulse shaping at the cost of nonlinear loss; 2, GVM would lead to asymmetric spectral broadening that can be utilized to better bridge different gain crystals; 3, the blue and red shifts are optimized at different wave-vector mismatches, and it is generally advantageous for mode-locking to utilize the red detuned side when *GVM*·Δ*k* <0 and the blue detuned side when *GVM*·Δ*k* >0; 4, nonlinear saturation leads to ripple generation and pulse distortion, but GVM in the pulse compression regime (*GVD*·Δ*k* <0) can counteract its effect and stabilize the mode-locking at a higher energy level.

The design rules presented in this paper can be applied to Mamyshev mode-locking of other solid-state platforms such as thin disk lasers that promise ever higher pulse energy and average power [36-38]. Recently, such cascaded quadratic nonlinearity in PPLN waveguide has been utilized to demonstrate supercontinuum generation with a spectrum that spans multiple octaves [39]. With such extensive spectral broadening and sophisticated active multipass geometry [40, 41] to further enhance the roundtrip gain, ultrabroadband solid-state Mamyshev oscillator that bridges even more spectrally separated gain crystals may become attainable. Finally, the control capability of both sign and magnitude of the effective Kerr nonlinearity provides a route to extend the application of Mamyshev oscillator in longer wavelengths.

## Funding

This work has been supported by the Office of Naval Research (ONR) under award number N00014-19-1-2251.

## Acknowledgment

The authors gratefully acknowledge helpful instructions of COMSOL software from Dr. Kunpeng Jia.